\documentclass[aps,prl,showpacs,twocolumn]{revtex4}
\usepackage{amsmath,amsfonts,amssymb,bm}				
\allowdisplaybreaks[1]
\usepackage[mathscr]{eucal}
\usepackage{nicefrac}														
\usepackage[T1]{fontenc}												
\usepackage{dsfont}														
\usepackage{graphicx}
\usepackage{subfigure}
\usepackage{float}															
\usepackage{wrapfig}
\usepackage{sidecap}														
\usepackage{placeins}

\usepackage{color}
\definecolor{mblue}{rgb}{0,0,0.91}
\newcommand{\mblue}{\color{mblue}}

\newcommand{\bra}[1]{\langle #1 |}
\newcommand{\ket}[1]{|#1\rangle}
\newcommand{\ketbra}[2]{|#1 \rangle\langle #2|}
\newcommand{\braket}[2]{\langle #1 |#2 \rangle}

\begin{document}
\title{Superradiant Phase Transitions and the Standard Description of Circuit QED}
\author{Oliver Viehmann,$^1$ Jan von Delft,$^1$ Florian Marquardt$^2$}
\affiliation{$^1$Physics Department,
             Arnold Sommerfeld Center for Theoretical Physics,
             and Center for NanoScience,\\
             Ludwig-Maximilians-Universit\"at,
             Theresienstra{\ss}e 37,
             80333 M\"unchen, Germany}
\affiliation{$^2$Institut for Theoretical Physics, Universit\"at Erlangen-N\"urnberg, Staudtstra\ss e 7, 91058 Erlangen, Germany}

\begin{abstract}
We investigate the equilibrium behaviour of a superconducting circuit QED system containing a large number of artificial atoms. It is shown that the currently accepted standard description of circuit QED via an effective model fails in an important aspect: it predicts the possibility of a superradiant quantum phase transition, even though a full microscopic treatment reveals that a no-go theorem for such phase transitions known from cavity QED applies to circuit QED systems as well. We generalize the no-go theorem to the case of (artificial) atoms with many energy levels and thus make it more applicable for realistic cavity or circuit QED systems. 
\end{abstract}

\maketitle

Recent years have seen rapid progress in fabrication and experimental control of superconducting circuit QED systems, in which a steadily increasing number of artificial atoms interact with microwaves~\cite{Wallraff2004,Fink2009,Neeley2010,DiCarlo2010}. These developments set the stage to study collective phenomena in circuit QED. An interesting question in that context is whether a system with many artificial atoms undergoes an equilibrium quantum phase transition as the coupling of artificial atoms and electromagnetic field is increased. Quantum phase transitions of this type have been intensely discussed for cavity QED systems~\cite{Raimond2001,Hepp1973,Wang1973,Rzazewski1975,Emary2003,Nataf2010b} and are known as superradiant phase transitions (SPTs)~\cite{Hepp1973}. However, in cavity QED systems with electric dipole coupling their existence is doubted due to a no-go theorem~\cite{Rzazewski1975}. Recently, it has been claimed that SPTs \emph{are} possible in the closely related circuit QED systems with capacitive coupling~\cite{Chen2007,Lambert2009,Nataf2010b}. This would imply that the no-go theorem of cavity QED does not apply and challenges the well-established analogy of circuit and \mbox{cavity QED.} 

Here, we show in a full microscopic analysis that circuit QED systems are also subject to the no-go theorem. We argue that such an analysis is necessary since the standard description of circuit QED systems by an effective model (EM) is deficient in the regime considered here. A toy model is used to illustrate this failure of an EM. Finally, we close a possible loophole of the no-go theorem by generalizing it from two-level to multi-level (artificial) atoms. Thus, our work restores the analogy of circuit and cavity QED and rules out SPTs in these systems under realistic conditions that have not been covered before.

\begin{figure}
	\centering
	\subfigure[]{\includegraphics[trim = 0pt 0pt 0pt 0pt, width=0.235\textwidth]{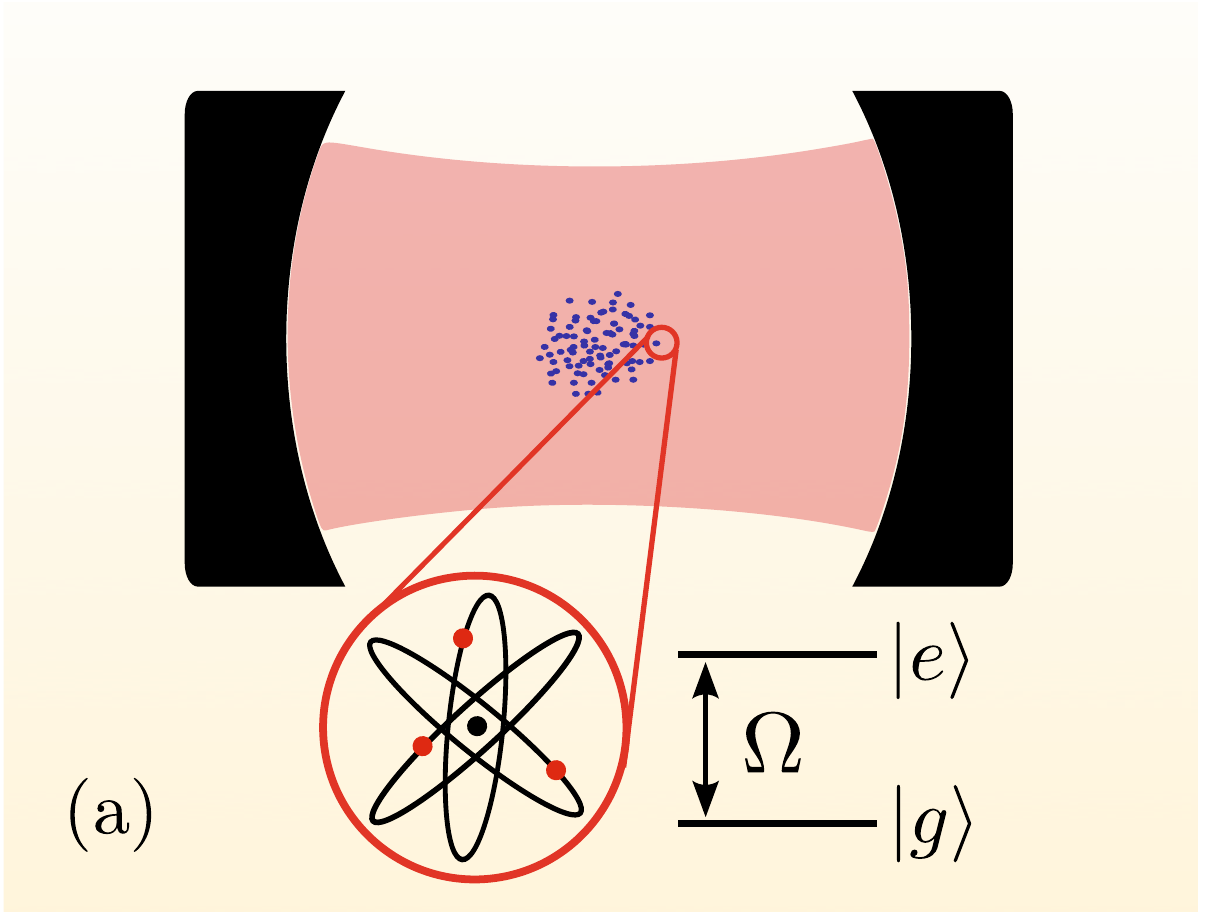}}\hspace{0cm}
	\vspace{0cm}
	\subfigure[]{\includegraphics[width=0.235\textwidth]{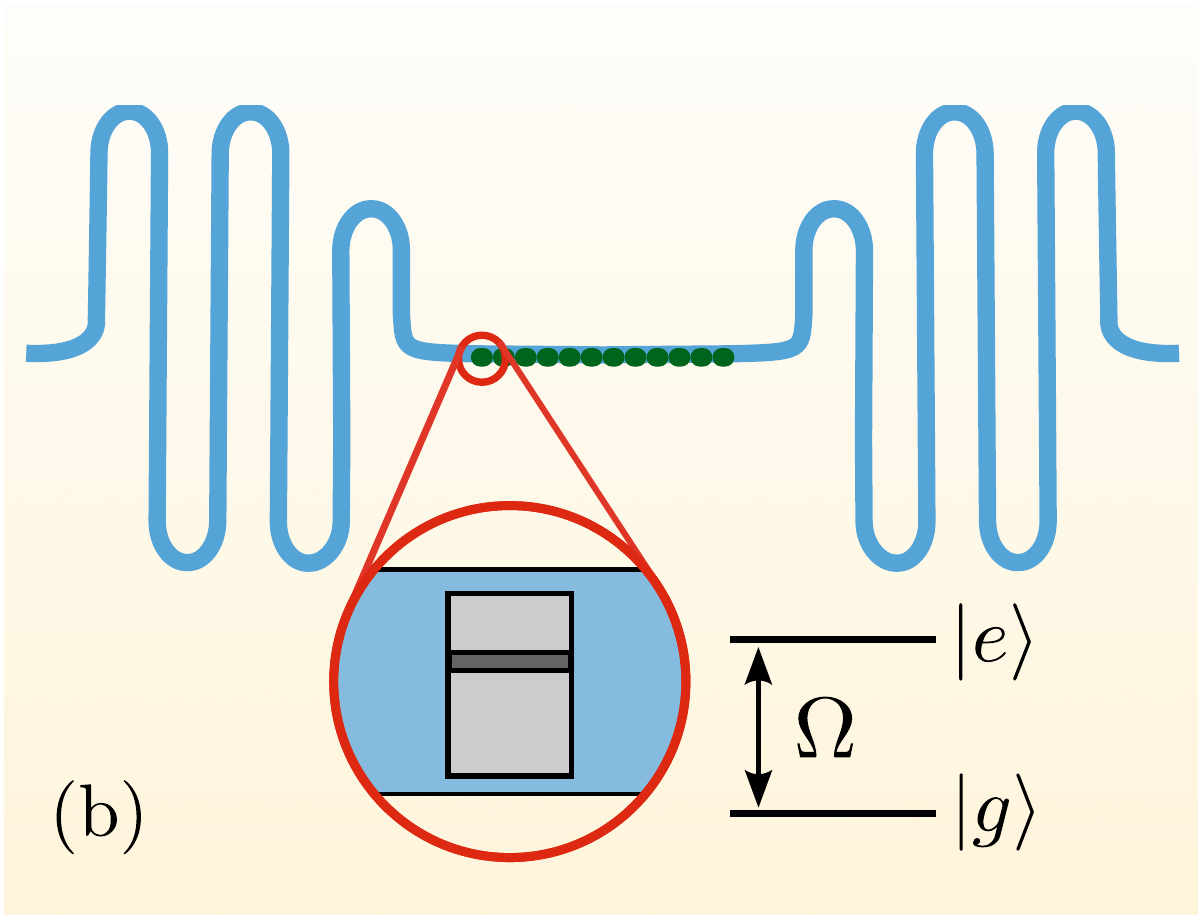}}
	\vspace{-0.3cm}
	\caption{Cavity QED system with $N$ atoms (a) and circuit QED system with $N$ Cooper-pair boxes as artificial atoms~(b).}
	\label{fig:CavityCircuitQED}
	\end{figure}

\paragraph*{The Dicke Hamiltonian in cavity and circuit QED.---}Both circuit QED systems and cavity QED systems with $N$ (artificial) atoms (Fig.\ \ref{fig:CavityCircuitQED})
are often described by the Dicke Hamiltonian~\cite{Dicke1954} ($\hbar = 1$)
	\begin{align*}
	\begin{split}
	\mathcal{H}_{\rm D}= \omega a^\dagger a  +   \dfrac{\Omega}{2} \! \sum_{k=1}^{N}{\! \sigma_z^k} \! + \!
	\dfrac{ \lambda}{\sqrt{N}} \!	\sum_{k=1}^N { \! \sigma_x^k (a^\dagger \! + \! a)}	 +  \kappa (a^\dagger \!+ \! a )^2 .
	\end{split}
	\end{align*}
The (artificial) atoms are treated as two-level systems with energy splitting $\Omega$ between ground state $\ket{g}_k = \left( \begin{smallmatrix} 0 \\ 1 \end{smallmatrix} \right)_k$ and excited state $\ket{e}_k = \left( \begin{smallmatrix} 1 \\ 0 \end{smallmatrix} \right)_k$ ($\sigma_x^k , \sigma_z^k$ are Pauli matrices). In the case of circuit QED, we assume Cooper-pair boxes as artificial atoms, which justifies the two-level approxiation. Our main results, though, hold for arbitrary charge-based artificial atoms~(capacitive coupling)~\cite{Clarke2008}. Further, $a^\dagger$ generates a photon of energy $\omega$. Matter and field couple with a strength $\lambda$. The $\kappa$-term, often neglected in other contexts, will become crucial below. In cavity QED, $\mathcal{H}_{\rm D}$ is obtained from minimal coupling of atoms and electromagnetic field. For a single atom (with $n$ electrons) at a fixed position, the Hamiltonian reads 
	\begin{align}
	\mathcal{H}_{\rm cav}^0= \sum_{i=1}^n{\dfrac{(\mathbf{p}_i-e\mathbf{A}(\mathbf{r}_i))^2}{2m} +V_{\rm int}(\mathbf{r}_1, \ldots, \mathbf{r}_n) }.
	\end{align}
The $\mathbf{pA}$- and $\mathbf{A}^2$-terms in the analog $N$-atom Hamiltonian yield the $\lambda$- and $\kappa$-term in $\mathcal{H}_{\rm D}$, respectively. In circuit QED, $\mathcal{H}_{\rm D}$ arises from a widely used EM for a charge\-based artificial atom in a transmission line resonator~\cite{Blais2004}, 
	\begin{align*}
	\begin{split}
	\mathcal{H}_{\rm cir}^0 \! = 4E_{\rm C} \! \sum_{\nu}{\!( \nu \!-\! \bar{\nu} )^2\ket{\nu}\bra{\nu}} 
	-	\dfrac{E_{\rm J}}{2}\! \sum_{\nu}{( \ket{\nu \! + \!1 }\bra{\nu} \!+\! 	{\rm H.c.} ) }  .
	\end{split}
	\end{align*}
Here, $\nu$ counts the excess Cooper pairs on the island, $E_{\rm J}$ and $E_{\rm C} = e^2/(2(C_{\rm G}+C_{\rm J}))$ are the Josephson energy and the charging energy of the Cooper-pair box, $C_{\rm G}$ and $C_{\rm J}$ are the coupling capacitance and the capacitance of the Josephson junction. Moreover, $\bar{\nu}=C_{\rm G} (V_{\rm G}+ \mathcal{V})/2e$, where $V_{\rm G}$ is an external gate voltage and $\mathcal{V}$ the quantum voltage due to the electromagnetic field in the resonator. The Cooper-pair box is assumed to be operated at its degeneracy point~\cite{Blais2004}. As it is described by macroscopic quantities (like $E_{\rm C}$) and only a single degree of freedom ($\nu$), $\mathcal{H}_0^{\rm cir}$ is an EM for a Cooper-pair box in a transmission line. Starting either from $\mathcal{H}_{\rm cav}^0$ or from $\mathcal{H}_{\rm cir}^0$, one obtains $\mathcal{H}_{\rm D}$ using the following approximations: The $N$ (artificial) atoms are treated as identical, noninteracting two-level systems with ground states $\ket{g}$ and excited states $\ket{e}$ which are strongly localized compared to the wavelength of the single considered field mode (i.e., $\mathbf{A}(\mathbf{r}_i^{k})\approx \mathbf{A} = A_0 \boldsymbol{\epsilon}(a^\dagger + a)$ and $\mathcal{V}(\mathbf{r}^{k}) \approx \mathcal{V} =   V_0 (a^\dagger+a)$).

\begin{figure}
	\centering
	\subfigure[]{\includegraphics[width=0.32\textwidth]{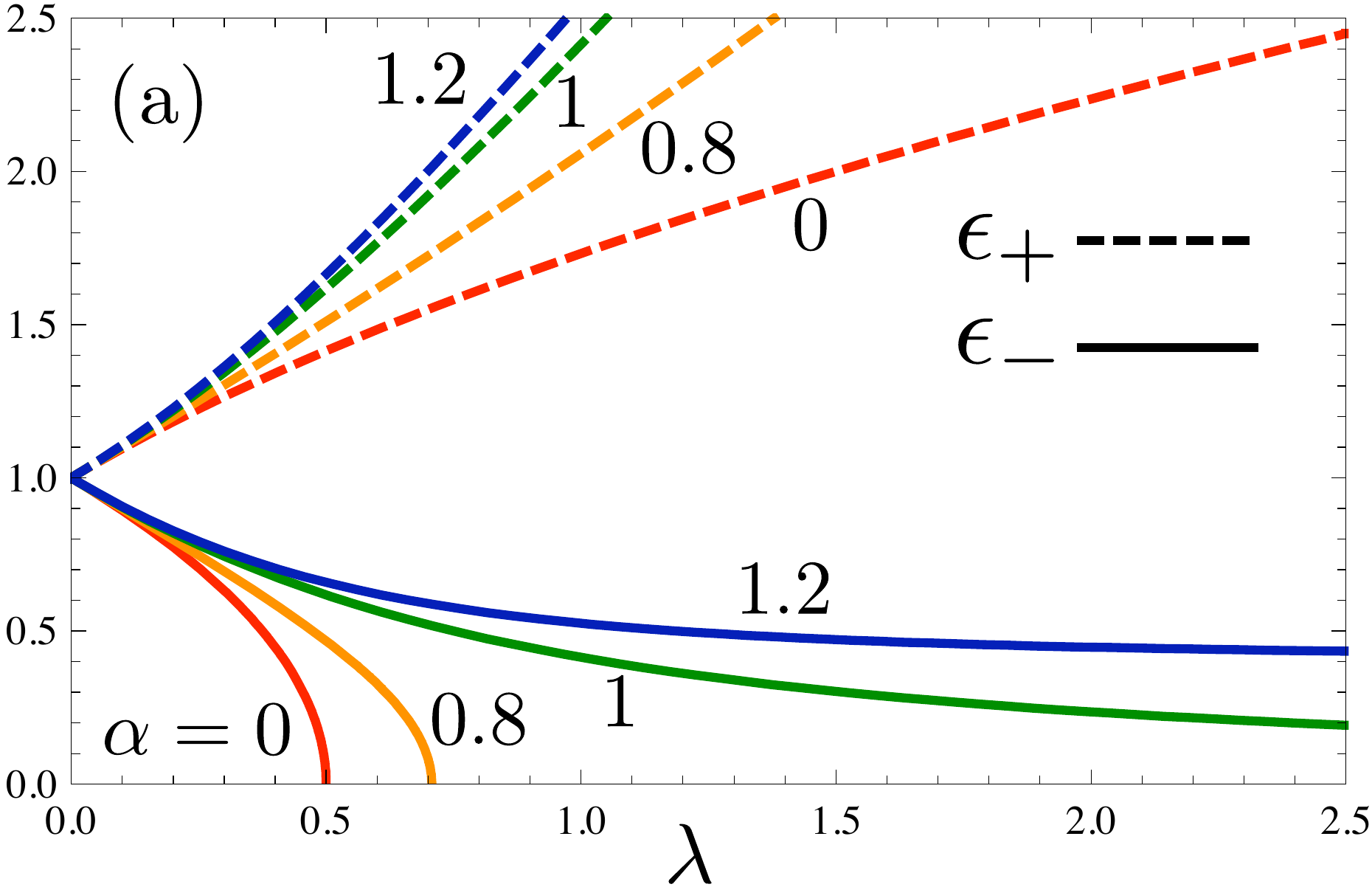}}
	\hspace{0.2cm}\vspace{0cm}	
	\subfigure[]{\raisebox{0.4cm}[0pt]{\includegraphics[width=0.14\textwidth]{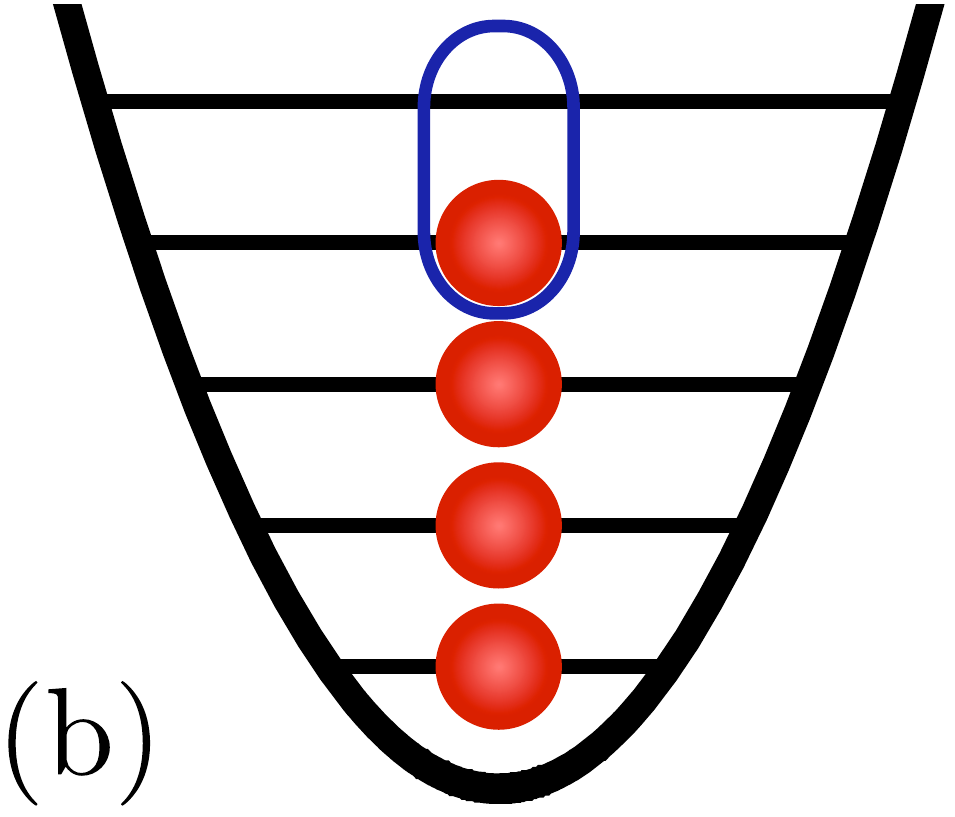}}}\vspace{-0.3cm}
	\caption{
	(a) Excitation energies $\epsilon_{+}$ and $\epsilon_{-}$ of the Dicke Hamiltonian 
	$\mathcal{H}_{\rm D}$ versus coupling $\lambda$ (in units of $\omega = \Omega$), for $\alpha = \kappa \Omega / \lambda^2 = 0, 0.8, 1, 1.2$. 
	For $\alpha = 0$, $\epsilon_{-}$  vanishes at $\lambda= 0.5$, thus signaling a SPT.
	Only $\alpha \geq 1$ is compatible with the TRK sum rule. For these $\alpha$, $\epsilon_{-} \rightarrow \sqrt{1-1/\alpha}$ and remains finite for all $\lambda$. 
	The excitations $\epsilon_{\pm} (\lambda)$
	of $\mathcal{H}_{\rm tm}$ correspond to $\alpha = 1$.
	(b)~Toy model of an (artificial) atom. The blue line indicates the degree of freedom in the simplified effective model. }
	\label{fig:ToyModel}
	\end{figure}

\paragraph*{Superradiant phase transitions and no-go theorem.---\hspace{-0.3cm}} 
In the limit $N \rightarrow \infty$, $\mathcal{H}_{\rm D}$ undergoes a second order quantum phase transition at a critical coupling strength~\cite{Hepp1973,Wang1973,Rzazewski1975}
	\begin{align}\label{eq:crit_coupling}
	\lambda_{\rm c}^2 = \dfrac{\omega \Omega}{4}\, \big( 1 + \dfrac{4 \kappa}{\omega} \big). 
	\end{align}
This phase transition was discovered by Hepp and Lieb for $\mathcal{H}_{\rm D}$ with $\kappa = 0$  and termed SPT~\cite{Hepp1973}; see~\cite{Emary2003} for recent studies. At $\lambda_{\rm c}$, the atoms polarize spontaneously, $\langle \sum_k \sigma_z^k \rangle / N \neq -1$, and a finite macroscopic photon occupation arises
, $\langle a^\dagger a \rangle / N \neq 0 $. This quantum critical point is signaled by a gapless excitation (Fig.\ \ref{fig:ToyModel}(a)). 

In cavity QED systems, however, $\lambda_{\rm c}$ cannot be reached if the $\kappa$-term is not neglected~\cite{Rzazewski1975}. That is because $\lambda$ and $\kappa$ are not independent of each other. Let us define a parameter $\alpha$ via $\kappa = \alpha \lambda^2/\Omega$. Then Eq.\ \eqref{eq:crit_coupling} becomes $\lambda_{\rm c}^2(1-\alpha) = \omega \Omega/4$, and criticality requires $\alpha < 1$. With $A_0=1/\sqrt{2 \epsilon_0 \omega V}$ ($V$ is the volume of the cavity) one finds
	\begin{align}\label{eq:CavQED_parameters}
	\lambda_{\rm cav} = \dfrac{\Omega |\boldsymbol{\epsilon} \cdot \mathbf{d}|}{\sqrt{2 \epsilon_0 \omega}} \sqrt{\dfrac{N}{V}}, \quad
	\kappa_{\rm cav} = \dfrac{n}{2 \epsilon_0 \omega } \dfrac{e^2}{2m} \dfrac{N}{V},
	\end{align}
where $\mathbf{d} = \bra{g} e \sum_{i=1}^n{\mathbf{r}_i }\ket{e}$, and $\alpha_{\rm cav} \Omega |\boldsymbol{\epsilon} \cdot \mathbf{d}|^2 = ne^2/2m$. But the Thomas-Reiche-Kuhn sum rule (TRK)
	\begin{align}\label{eq:TRKcavity}
	\sum_l{ (E_l-E_g)\;| \boldsymbol{\epsilon}\cdot \bra{g}   e \sum_{i=1}^n { \mathbf{r}_i} \ket{l} |^2} = n \frac{ e^2}{2 m}
	\end{align} 
for the Hamiltonian $H^0= \sum_{i=1}^n{ \!\mathbf{p}_i^{2}/2m \!+ \! V_{\rm int}(\mathbf{r}_1, \ldots, \mathbf{r}_n) }$ of an uncoupled atom with spectrum $\{E_l, \ket{l} \}$ implies $\Omega |\boldsymbol{\epsilon} \cdot \mathbf{d}|^2 \leq n e^2/2m$, consequently $\alpha_{\rm cav} \geq 1$. This is known as the no-go theorem for SPTs~\cite{Rzazewski1975,Nataf2010b}. Notice that $\alpha_{\rm cav}$ determines how strongly $\Omega |\boldsymbol{\epsilon} \cdot \mathbf{d}|^2$ exhausts the TRK. We remark that a direct dipole-dipole coupling between atoms (omitted here) can lead to a ferroelectric phase transition, which however occurs only at very high atomic densities \cite{Keeling2007}.

Surprisingly, the no-go theorem was recently argued not to apply in circuit QED~\cite{Nataf2010b}. Indeed, the standard EM of circuit QED yields
	\begin{align}
	\lambda_{\rm cir}= \dfrac{e C_{\rm G}}{C_{\rm G} \! +\! C_{\rm J}} \sqrt{\dfrac{\omega N}{Lc}}, \quad 
	\kappa_{\rm cir}= \dfrac{C_{\rm G}^2}{2(C_{\rm G}\! +\! C_{\rm J})} \dfrac{\omega N}{Lc},
	\end{align}
where $L$ denotes the length of the transmission line resonator, $c$ its capacitance per unit length, and we have used $V_0=(\omega/Lc)^{1/2}$~\cite{Blais2004}. Here $\alpha_{\rm cir} = E_{\rm J}/4E_{\rm C} <1$ is easily possible \cite{Wallraff2004}. According to this argument, a SPT should be observable in a circuit QED system.

\paragraph*{Effective models and superradiant phase transitions.---}
The EM has proved to be a very successful description of circuit QED whose predictions have been confirmed in numerous experiments. However, the circuit QED setups operated so far contained only few artificial atoms. It is not obvious that an EM provides also a good description of circuit QED systems with $N \gg 1$ atoms and, thus, a proper starting point to study SPTs in circuit QED. We now present a toy model illustrating how an EM similar to the one in circuit QED erroneously predicts a SPT.

The toy model consists of $N$ harmonic oscillator potentials with frequency $\Omega$, each trapping $n$ noninteracting fermions of mass $m$ and charge $e$, which all couple to a bosonic mode with frequency $\omega$ (Fig.\ \ref{fig:ToyModel}(b)). 
This toy model can be viewed as a very simplified description of (artificial) atoms with $n$ microscopic constituents inside a resonator. It is governed by the Hamiltonian
	\begin{align}\label{eq:toymodel}
	\mathcal{H}_{\rm tm} = \omega a^\dagger a   +   \sum_{k=1}^N{ 	\sum_{i=1}^n{\dfrac{(p_i^k-eA)^2}{2m} + \dfrac{m\Omega^2 (x_i^k)^2}{2} }},
	\end{align}
where we assume again $A(x_i^k) \approx A = A_0 (a^\dagger \! + a)$. Since $A$ couples only to the center of mass coordinate of the $k$th harmonic oscillator,  $\mathcal{H}^{\rm }$ can be diagonalized exactly~(\cite{Supp}): 
	\begin{gather}
	\mathcal{H}_{\rm tm} = \epsilon_{\pm} (a^\dagger_{\pm} a_{\pm} + \dfrac{1}{2} )  + \sum_{i=1}^{nN-1}{\Omega (b_i^\dagger b_i + \dfrac{1}{2}) }, \\
	2 \epsilon_{\pm}^2(\lambda) =  
	\omega^2 \! \! + 4 \kappa \omega \! + \! \Omega^2 
	\pm \sqrt{ (\omega^2 \! \! +  4\kappa \omega \! - \! \Omega^2 )^2 + 16 \lambda^2 \omega \Omega  }. \notag
	\end{gather}
Here, $a_\pm^\dagger$ generate  excitations that mix photon field with collective center of mass motion, the $b_i^\dagger$ excite the remaining degrees of freedom, $\lambda = A_0 \Omega d \sqrt{N}$ and $\kappa=\lambda^2 / \Omega $. As $d = \langle n | e x | n-1 \rangle = e\sqrt{n/2m\Omega}$, the TRK is exhausted. Note that  $\epsilon_{\pm}(\lambda)$ are also the relevant excitation energies of $\mathcal{H}_{\rm D}$ for $N \rightarrow \infty$, as can be shown using methods of Ref.\ \cite{Emary2003} (see \cite{Supp}), and demanding $\epsilon_- = 0$ yields Eq.\ \eqref{eq:crit_coupling}. One sees that $\epsilon_{\pm}(\lambda)$ is real and nonzero for all $\lambda $ and that the ground state energy is an analytic function of $\lambda$ (Fig.\ \ref{fig:ToyModel}(a)). Hence, no quantum phase transition is possible.

Let us now consider an EM for the toy model. Similar to the standard EM of circuit QED, we focus on the fermion with the highest energy in the $k$th harmonic oscillator and treat it as two-level system with $\ket{g_k} = \ket{n-1}_k$ and $\ket{e_k} = \ket{n}_k$ (Fig.\ \ref{fig:ToyModel}(b)). Accounting only for one fermion per `atom' yields a Dicke Hamiltonian with $\lambda_{\rm em} =\lambda$ and $\kappa_{\rm em} = \kappa/n = N e^2 A_0^2/2m$. Crucially, only $\lambda_{\rm em}$ depends on $n$. This allows $\lambda_{\rm em}$ to be increased at constant $\kappa_{\rm em}$, therefore $\alpha_{\rm em} = 1/n$ can be $<1$ and a SPT is possible. This failure of the EM can be interpreted as follows. The relation $\lambda = \lambda_{\rm em} \propto d \propto \sqrt{n}$ reveals that the coupling of an `atom' to the bosonic mode is fully captured by the EM and grows with atom size $n$. However, in a proper description of the system, increasing the coupling by increasing $n$ unavoidably also increases $\kappa$ in proportion to $n$: \textit{all} fermions of all atoms couple to the bosonic mode and each causes an $A^2$-term. This is lost in the EM with only one degree of freedom per atom. Interestingly, $\alpha_{\rm em}<1$ only if $n>1$, i.e., as long as the effective description actually neglects degrees of freedom.

\paragraph*{Microscopic description of circuit QED.---\hspace{-0.3cm}}This example suggests not to rely on the standard description for investigating SPTs in circuit QED. Although the dipole coupling of field and qubit states might be fully represented by $\lambda_{\rm cir}$, $\kappa_{\rm cir}$ could still underestimate the $A^2$-terms of all charged particles in the Cooper-pair boxes. Instead, let us describe a circuit QED system with $N$ artificial atoms by a minimal-coupling Hamiltonian that takes all microscopic degrees of freedom into account:
	\begin{align*}
	\mathcal{H}_{\rm mic} =  \omega a^\dagger a + \sum_{k=1}^N{ 
	\sum_{i=1}^{n_k}{ \dfrac{(\mathbf{p}_i^{k} - q_i^k \mathbf{A})^2}{2 m_i^k} + V_{\rm int}(\mathbf{r}_1^{k}, \ldots , \mathbf{r}_{n_k}^{k})}.
	} 
	\end{align*}
As we allow arbitrary charges $q_i^k$ and masses $m_i^k$ and an arbitrary interaction potential $V_{\rm int}$ of the $n_k$ constituents of the $k$th artificial atom, $\mathcal{H}_{\rm mic}$ captures the coupling of $N$ arbitrary (but mutually noninteracting) objects to the electromagnetic field in the most general way. We subject it to the same approximations that led from $\mathcal{H}^0_{\rm cir}$, the EM of circuit QED, to $\mathcal{H}_{\rm D}$. For identical artificial atoms $\{ n_k, q_i^k, m_i^k \} \rightarrow \{ n, q_i, m_i \}$. The Hamiltonian of an uncoupled artificial atom then reads $H_{\rm mic}^0= \sum_{i=1}^n{\mathbf{p}_i^{2}/2m_i +V_{\rm int}(\mathbf{r}_1, \ldots, \mathbf{r}_n) }$. Notice that the qubit states $\ket{g}$ and $\ket{e}$ of the artificial atom, which in the standard EM are superpositions of the charge states $\ket{\nu}$, are among the eigenstates $\{ \ket{l} \}$ of $H_{\rm mic}^0$. By expanding $\mathcal{H}_{\rm mic}$ in the $\{ \ket{g}_k, \ket{e}_k \}$-basis and taking $\mathbf{A}(\mathbf{r}_i^{k}) \approx \mathbf{A}$, one obtains the Dicke Hamiltonian $\mathcal{H}_{\rm D}$ with
	\begin{align}\label{eq:CirQED_mic_parameters}
	\lambda_{\rm cir}^{\rm mic}= \dfrac{\Omega |\boldsymbol{\epsilon} \cdot \mathbf{d}|}{\sqrt{2 \epsilon_0 \omega}} \sqrt{\dfrac{N}{V}}, \quad 
	\kappa_{\rm cir}^{\rm mic} = \dfrac{1}{2 \epsilon_0 \omega } \!  \Big( \sum_{i=1}^n{\dfrac{q_i^2}{2m_i}} \Big)  \!\dfrac{N}{V},
	\end{align}
and $\mathbf{d} = \bra{e} \sum_{i=1}^n{q_i \mathbf{r}_i }\ket{g}$. This microscopic description of circuit QED facilitates the same line of argument which in Ref.\ \cite{Rzazewski1975} allowed the conclusion that there is no SPT in cavity QED: Criticality (Eq.\ \eqref{eq:crit_coupling}) requires $\Omega |\boldsymbol{\epsilon} \cdot \mathbf{d}|^2 > \sum_{i=1}^n{q_i^2/2m_i}$, which is ruled out by TRK for $H_{\rm mic}^0$,
	\begin{align}\label{eq:TRKcircuit}
	\sum_{l}{(E_l-E_g) |\boldsymbol{\epsilon} \cdot \bra{g}   \sum_{i=1}^n { q_i \mathbf{r}_i} \ket{l} |^2} = \sum_{i=1}^n{\frac{ q_i^2}{2 m_i}}.
	\end{align}
Hence, \emph{the no-go theorem of cavity QED applies to circuit QED as well}. This result confirms the analogy of cavity and circuit QED also with respect to SPTs. It has been obtained under the same approximations that led from the standard description of circuit QED, $\mathcal{H}_{\rm cir}^0$, to $\mathcal{H}_{\rm D}$ with $\lambda_{\rm cir}$ and $\kappa_{\rm cir}$. The discrepancy of the predictions of the microscopic and the standard description of circuit QED thus shows the limitations of the validity of the latter. This might be important for future circuit QED architectures with many artificial atoms in general, even for applications not related to SPTs. We emphasize, though, that our conclusion neither forbids SPTs in circuit QED systems with inductively coupling flux qubits~\cite{Nataf2010a}, nor is it at odds with the great success of the standard description for \emph{few}-atom systems: there, the deficiency of $\kappa_{\rm cir}$ does not manifest itself qualitatively as the $\kappa$-term in $\mathcal{H}_{\rm D}$ mimics slightly renormalized system parameters $\tilde{\omega}$ and $\tilde{\lambda}$.

\paragraph*{Possible loophole in the no-go theorem.---\hspace{-0.3cm}}Although the two-level approximation for the anharmonic spectrum of (artificial) atoms is well justified in many cases, one might argue that higher levels should be taken into account in this context. Indeed, a SPT does not require $\Omega \approx \omega$, and thereby does not single out a particular atomic transition. 

For a more profound reason for dropping the two-level assumption, consider the elementary question of how a resonator's frequency $\omega$ is shifted by the presence of $N$  mutually noninteracting atoms. This situation is described by $\mathcal{H}_{\rm mic}$. It can be rewritten as $\mathcal{H}_{\rm mic} = \omega a^\dagger a + \sum_{k=1}^N{ \left( H_{\rm mic}^k + \mathcal{H}_{pA}^k + \mathcal{H}_{A^2}^k \right) } $, where $\mathcal{H}_{pA}^k$ and $\mathcal{H}_{A^2}^k $ are the $\mathbf{pA}$- and the $\mathbf{A}^2$-terms due to the $k$th atom (\cite{Supp}). Let us perturbatively calculate the frequency shift $\delta \omega = \delta \omega _{pA}+ \delta \omega _{A^2}$ caused by both $\sum \mathcal{H}_{pA}^k$ and $\sum \mathcal{H}^k_{A^2}$ (\cite{Supp}). To this end, take $\omega \ll \Omega_{m}^k$ for all $m,k$, where $\Omega_{m}^k$ is the $m$th excitation energy of $H_{\rm mic}^k$. Remarkably, it turns out (\cite{Supp}) that $\delta \omega_{pA}$ $(<0)$ and $\delta \omega_{A^2}$ $(>0)$ cancel almost exactly by virtue of the TRK, and the total frequency shift $\delta \omega$ is small, $\delta \omega \sim (\omega/\Omega_{m}^k)^2$. As a SPT corresponds to $\delta \omega = -\omega$, the significance of both $\mathbf{pA}$- and $\mathbf{A}^2$-terms for the existence of a SPT becomes clear. The $\mathbf{pA}$-terms cause a strong negative frequency shift and favor a SPT, the $\mathbf{A}^2$-terms do the opposite. This means, most crucially, that one must not truncate $\mathbf{pA}$- and $\mathbf{A}^2$-terms in an unbalanced way for assessing the possibility of a SPT by an approximate Hamiltonian. Dropping the $\mathbf{A}^2$-terms in $\mathcal{H}_{\rm D}$ ($\kappa = 0$) leads to the prediction of a SPT. In contrast, $\mathcal{H}_{\rm D}$ with nonzero $\kappa$ fully incorporates the $\mathbf{A}^2$-terms of $\mathcal{H}_{\rm mic}$. But, due to the two-level approximation, it allows for only one matrix element of the $\mathbf{pA}$-terms per atom, thereby possibly underestimating the tendency towards a SPT. To exclude SPTs in cavity and circuit QED, a generalization of the no-go theorem to (artificial) atoms with more than two energy levels is necessary.

\begin{figure}
	\hspace{0cm}
	\includegraphics[width=0.37\textwidth]{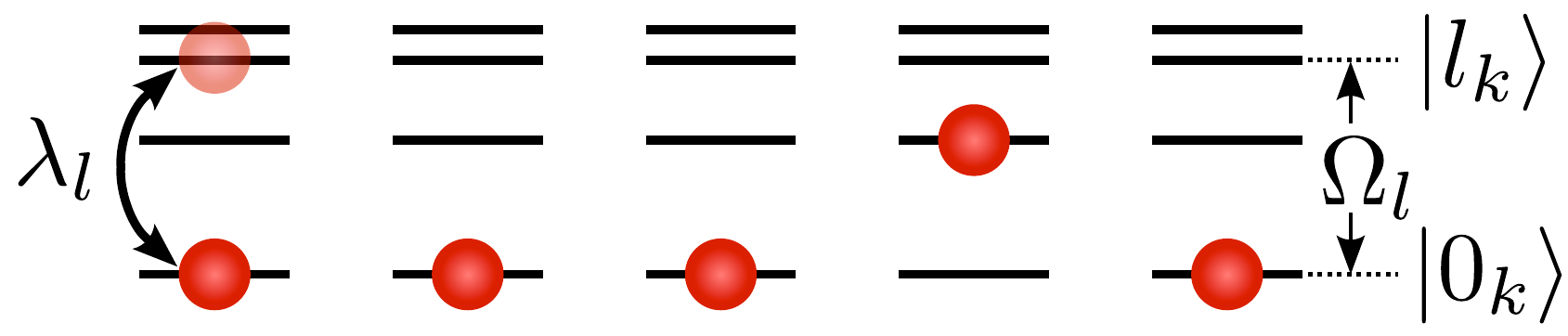}
	\vspace{-0.3cm}
	\caption{Situation of the generalized no-go theorem. Many multi-level (artificial) atoms couple to the photon field. 
	Transitions between excited atomic states are irrelevant for the low-energy spectrum of the system. }
	\label{fig:HgD}
	\end{figure}

\paragraph*{Generalized no-go theorem.---\hspace{-0.3cm}}Let us consider $N \rightarrow \infty$ identical atoms couped to a field mode with frequency $\omega$. The atomic Hamiltonians $H_{\rm mic}^k$ may have an arbitrary spectrum $\{ \Omega_l , \ket{l_k}=\ket{l}_k \}$, with $\Omega_0 = 0$ and $\mu$ excited states (Fig.\ \ref{fig:HgD}). 
	With $d_{l,l^\prime} = \boldsymbol{\epsilon} \cdot \bra{l} \sum_{i=1}^n{q_i \mathbf{r}_i} \ket{l^\prime}$, the full Hamiltonian of the system reads 
	\begin{align}\label{eq:Hmic_mt}
	\mathcal{H}_{\rm mic} =\;& \omega a^\dagger a + \kappa (a^\dagger + a)^2 +
	\sum_{k=1}^N
	\sum_{l,l^\prime =0}^{\mu} \Big(
	\Omega_l \delta_{l,l^\prime} \ketbra{l_k}{l_k} \notag \\
	&+ i A_0 (\Omega_{l^\prime} - \Omega_{l} ) d_{l,l^\prime}(a^\dagger + a ) \ketbra{l_k}{l^{\prime}_k} \Big).
	\end{align}
We now follow a strategy similar to that of Refs.~\cite{Emary2003}: We derive a \emph{generalized Dicke Hamiltonian} $\mathcal{H}_{\rm gD}$ having the same low-energy spectrum as $\mathcal{H}_{\rm mic}$ for a small density of atoms, $N/V \simeq 0$, using $A_0 \propto V^{-1/2}$ as small parameter. We then check whether $\mathcal{H}_{\rm gD}$ has a gapless excitation if the density is increased, which would signal a SPT and mark the breakdown of the analogy of $\mathcal{H}_{\rm gD}$ and $\mathcal{H}_{\rm mic}$. 

Expanding the eigenstates and -energies of $\mathcal{H}_{\rm mic}$ as $|\mathcal{E} \rangle \propto \sum_{s=0}^\infty A_0^s |\mathcal{E}_s \rangle$ and $\mathcal{E} \propto \sum_{ss'} A_0^{s+s'} \langle \mathcal{E}_s |  \mathcal{H}_{\rm mic} | \mathcal{E}_{s'} \rangle$, we note that contributions from all $d_{l \neq 0, l' \neq 0}$ terms may be neglected: they are smaller than those retained by a factor of at least one power of $A_0$ (for $s+s' > 1$)  or $\xi/N \ll 1$ (for $s+s' \le 1$), where $\xi = \sum_k \sum_{l>0}| \langle l_k|\mathcal{E}_0\rangle|^2$ is the number of atomic excitations in $|\mathcal{E}_0 \rangle$, which is $\ll N$ for low-lying eigenstates (\cite{Supp}).  We thus define $\mathcal{H}_{\rm gD} $ by setting $d_{l \neq 0, l' \neq 0} \to 0 $ in $\mathcal{H}_{\rm mic}$. Up to a constant, we find (\cite{Supp})
	\begin{align}\label{eq:Hgd_mt}
	\mathcal{H}_{\rm gD} = \tilde{\omega} a^\dagger a +
	\sum_{l=1}^\mu{ \Omega_l b_l^\dagger b_l }+ \sum_{l=1}^{\mu}{\tilde{\lambda}_l(b_l^\dagger \!+ b_l)(a^\dagger \!+a) } ,
	\end{align}
by introducing $b_l^\dagger = \frac{1}{\sqrt{N}}\sum_{k=1}^N \ketbra{l_k}{0_k}$ as collective excitation, omitting the energy of the `dark' collective excitations (\cite{Supp}), and removing the $\kappa$-term by a Bogolyubov transformation yielding $\omega \rightarrow \tilde{\omega} = \sqrt{\omega^2 + 4 \kappa \omega}$ and $\lambda_l \rightarrow \tilde{\lambda_l} = \sqrt{\frac{\omega}{\tilde{\omega}}} \lambda_l$, with $\lambda_l = A_0 \Omega_l |d_{0,l}| \sqrt{N}$. In the limit of dilute excitations, the $b_l$ are bosonic, $[ b_l,b_{l^\prime}^\dagger ] = \delta_{l,l^\prime} $~\cite{Hopfield1958}. The system undergoes a SPT if an eigenfrequency $\epsilon_i$ of $\mathcal{H}_{\rm gD}$ can be pushed to zero by increasing the couplings $\lambda_l$. We cannot calculate the $\epsilon_i$'s explicitly, but we will show that the assumption $\epsilon_i=0$ contradicts the TRK. An $\epsilon_i$ solves the characteristic equation (\cite{Supp})
	\begin{align}\label{eq:char_eq_mt}
	 \Big( \prod_{l^\prime =1}^{\mu}{(\Omega_{l^\prime}^2\! - \epsilon^2)} \Big)   \Big( (\tilde{\omega}^2 \! -\epsilon^2) 
	 - 4 \tilde{\omega}\sum_{l =1}^{\mu}{\dfrac{ \Omega_{l} \tilde{\lambda}_{l}^2 }{\Omega_{l}^2 \!- \epsilon^2}} \Big)  = 0.
	\end{align}
If $\epsilon_i$ were zero, this would imply
	\begin{align}\label{eq:contradiction}
	\dfrac{\omega}{4N A_0^2} =  \sum_{l=1}^{\mu}{\Omega_l} |d_{0l}|^2 -\sum_{i=1}^{n}{\dfrac{q_i^2}{2m_i}} 
	\end{align}
and contradict the TRK for $H_{\rm mic}^0$ (Eq.\ \eqref{eq:TRKcircuit}), which ensures that the right side is negative even if the entire atomic spectrum is incorporated. This result is irrespective of the details of the atomic spectra. Note that for $\kappa = 0$, the negative term on the right side of Eq.\ \eqref{eq:contradiction} vanishes, and one recovers the SPT for critical couplings $\lambda_{l {\rm c}}$ with $\sum_{l=1}^\mu \lambda_{l {\rm c}}^2/\Omega_l = \omega/4$. This resembles Eq.\ \eqref{eq:crit_coupling} with $\kappa = 0$.

Experimental evidence for our conclusions could be gained by simply probing the shifted resonator frequency of a suitable circuit QED system. Consider a sample containing $N$ artificial atoms with $\lambda/\sqrt{N} = 2 \pi \times 120\text{MHz}$ and $\Omega/2\pi = \omega/2\pi = 3 \text{GHz}$. If $\alpha_{\rm cir} = E_{\rm J}/4E_{\rm C} = 0.1$, as predicted by the standard theory of circuit QED, the system should exhibit signatures of criticality for $N = 174$ (according to Eq.\ \eqref{eq:crit_coupling}), and its resonator frequency should be close to zero. But even if we assume $\alpha = 1$, the minimal value compatible with the TRK (that corresponds to ideal two-level atoms), we find the lowest excitation $\epsilon_-$ to be still at $\epsilon_-\approx 2\pi \times 2\text{GHz}$. We expect these phenomena to be insensitive to small fluctuations of the atomic parameters~\cite{Nataf2010a} and hence to be experimentally observable.

We thank S. M. Girvin, A. Wallraff, J. Fink, A. Blais, J. Siewert, D. Esteve, J. Keeling, P. Nataf, and C. Ciuti for discussions. Support by NIM, the Emmy-Noether program, and the SFB 631 of the DFG is gratefully acknowledged.


\pagebreak
\appendix*
\onecolumngrid
\addtolength{\textwidth}{-3.58cm}
\addtolength{\oddsidemargin}{1.45cm}
\setlength{\belowcaptionskip}{0cm}
\part{{\fontsize{14}{14}\selectfont \vspace*{25mm} Appendix}}

\vspace{5mm}
We provide intermediate steps for the derivation of some important statements and equations of the main text. For clarity, formulas contained in
the main text are typeset in blue.

\vspace{8mm}
\noindent
{\fontsize{11}{12}\selectfont \textbf{Thomas-Reiche-Kuhn sum rule. }} We derive the TRK \cite{TRK1925_app} for the Hamiltonian 
	\begin{align}
	{\mblue
	H_{\rm mic}^0= \sum_{i=1}^n{\dfrac{\mathbf{p}_i^{2}}{2m_i} +V_{\rm int}(\mathbf{r}_1, \ldots, \mathbf{r}_n) }
	},	
	\end{align}
yielding Eq.\ \eqref{eq:TRKcircuit} of the main text; Eq.\ \eqref{eq:TRKcavity} follows as a special case. The derivation of the TRK is based upon the identities
	\begin{align}\label{eq:commutator}
	\sum_{i=1}^n \dfrac{q_i^2}{2m_i} = - i \Big[ \boldsymbol{\epsilon} \cdot \sum_{i=1}^n q_i \mathbf{r}_i\,, \boldsymbol{\epsilon} \cdot \sum_{i^\prime = 1}^n \dfrac{q_{i^\prime} \mathbf{p}_{i^\prime}}{2m_{i^\prime}} \Big] , \qquad \sum_{i=1}^{n}{\dfrac{q_i \mathbf{p}_i }{ m_i}} = i \Big[ H_{\rm mic}^0 , \sum_{i=1}^{n}{q_i \mathbf{r}_i} \Big],
	\end{align}	
for a real unit vector $\boldsymbol{\epsilon}$. We denote the eigenspectrum of $H_{\rm mic}^0$ by $ \{ E_{l}, \ket{l} \}$. It comprises a ground state $\ket{g}$ of energy $E_g$. The TRK follows by combining the commutators of Eqs.~\eqref{eq:commutator}:
	\begin{subequations}
	\begin{align}
	{\mblue	
	\sum_{i=1}^n \dfrac{q_i^2}{2m_i}
	} &= \bra{g} \Big[ \boldsymbol{\epsilon} \cdot \sum_{i=1}^n q_i \mathbf{r}_i\,, \dfrac{\boldsymbol{\epsilon}}{2} \cdot \Big[  H_{\rm mic}^0, \sum_{i^\prime =1}^{n}{q_{i^\prime} \mathbf{r}_{i^\prime}}  \Big] \Big]\ket{g}\\
	&{\mblue = \sum_l (E_l-E_g) |  \boldsymbol{\epsilon} \cdot \bra{g} \sum_{i=1}^n q_i \mathbf{r}_i \ket{l}|^2}.
	\end{align}
	\label{eq:TRK}
	\end{subequations}	
\vspace{8mm}

\noindent
{\fontsize{11}{12}\selectfont \textbf{Diagonalization of $\mathcal{H}_{\rm D}$ and $\mathcal{H}_{\rm tm}$. }} It is demonstrated that the diagonalization of both the Dicke Hamiltonain $\mathcal{H}_{\rm D}$ for $N \rightarrow \infty$ and the Hamiltonian $\mathcal{H}_{\rm tm}$ describing the toy model can be reduced to the diagonalization of special cases of $\mathcal{H}_{\rm gD}$, which appears in the context of the generalized no-go theorem. The characteristic equation of $\mathcal{H}_{\rm gD}$, that will be derived in the last section of this appendix, is solvable for the special cases and yields the diagonal forms of $\mathcal{H}_{\rm D}$ and $\mathcal{H}_{\rm tm}$.

\vspace{5mm}
\noindent
{\fontsize{10}{12}\selectfont \textbf{Diagonalization of $\mathcal{H}_{\rm D}$. }} First, we focus on the Dicke Hamiltonian 
	\begin{subequations}	
	\begin{align}
	{\mblue
	\mathcal{H}_{\rm D}} & 
	{\mblue = \omega a^\dagger a   +   \dfrac{\Omega}{2}  \sum_{k=1}^{N}{ \sigma_z^k}  + 
	\dfrac{ \lambda}{\sqrt{N}} 	\sum_{k=1}^N {  \sigma_x^k (a^\dagger  +  a)}	 +  \kappa (a +  a^\dagger )^2
	}\\
	& =  \tilde{\omega} a^\dagger a  +   \dfrac{\Omega}{2}  \sum_{k=1}^{N}{ \sigma_z^k}  + 
	\dfrac{ \tilde{\lambda}}{\sqrt{N}} 	\sum_{k=1}^N {  \sigma_x^k (a^\dagger  +  a)} + C\label{eq:H_D_Bog}
	\end{align} 
	\end{subequations}
with $\tilde{\omega} = \sqrt{\omega^2 + 4 \kappa \omega}$, $\tilde{\lambda} = \sqrt{\omega/\tilde{\omega}} \lambda$, and $C=(\tilde{\omega} - \omega)/2$. The Hamiltonian \eqref{eq:H_D_Bog} was diagonalized by means of a Holstein-Primakoff transformation in Refs.~\cite{Emary2003_app}. We employ here a closely related approach developed in~\cite{Hopfield1958_app}, which is more convenient for a generalization beyond the two-level approximation and was also used in the derivation of $\mathcal{H}_{\rm gD}$. We drop $C$, set the energy of the atomic ground states to zero, introduce the operators
	\begin{align}
	a_k^\dagger = \ketbra{e_k}{g_k}, \qquad  b_{q_j}^\dagger = \dfrac{1}{\sqrt{N}} \sum_{k=1}^N e^{i q_j k} \ketbra{e_k}{g_k},
	\end{align}
where $q_j = 2\pi (j/N)$ and $j \in \{0,1,\ldots,N-1\}$, and obtain for $N \rightarrow \infty$
	\begin{subequations}
	\begin{align}
	\mathcal{H}_{\rm D}^\prime &= \tilde{\omega} a^\dagger a  + \Omega  \sum_{k=1}^{N}{ a_k^\dagger a_k}  +	\tilde{\lambda} 	(b_{q_0}^\dagger + b_{q_0})(a^\dagger  +  a) \\
	& = \tilde{\omega} a^\dagger a  +  \Omega \sum_{j=0}^{N-1}{ b_{q_j}^\dagger b_{q_j}}  +	\tilde{\lambda}  (b_{q_0}^\dagger + b_{q_0})	(a^\dagger  +  a).
	\end{align}
	\end{subequations}
In the limit of dilute excitations, that is applicable as long as the excitation energies of the system are finite, the $b_{q_j}$ obey bosonic commutation relations. Note that only the $j=0$ collective mode couples to the radiation field. The $j \neq 0 $ modes are `dark' and will be omitted in the following. We write $b$ instead of $b_{q_0}$ and arrive at 
	\begin{align}	
	\mathcal{H}_{\rm D}^{\prime \prime} = \tilde{\omega} a^\dagger a  +  \Omega b^\dagger b  +	\tilde{\lambda} 	 (b^\dagger + b)(a^\dagger  +  a),
	\end{align}
which corresponds to $\mathcal{H}_{\rm gD}$ (Eqs.\ \eqref{eq:Hgd_mt} and \eqref{eq:H_gD}) with $\mu = 1$. Later we will derive a characteristic equation for the eigenfrequencies of $\mathcal{H}_{\rm gD}$ (Eqs.\ \eqref{eq:char_eq_mt} and \eqref{eq:chareq}). For $\mu=1$ this equation has the solutions
	\begin{align}
	{\mblue 2 \epsilon_{\pm}^2 =  \omega^2  + 4 \kappa \omega +  \Omega^2 
	\pm \sqrt{ (\omega^2  +  4\kappa \omega -  \Omega^2 )^2 + 16 \lambda^2 \omega \Omega}
	}.
	\end{align}

\vspace{5mm}
\noindent
{\fontsize{10}{12}\selectfont \textbf{Diagonalization of $\mathcal{H}_{\rm tm}$. }} Now we consider $\mathcal{H}_{\rm tm}$ (Eq.\ \eqref{eq:toymodel}). The coupling of the electromagnetic field and a single harmonic oscillator `atom' is described by
	\begin{align}
	\mathcal{H}_{\rm tm}^0 = \sum_{i=1}^{n}{\dfrac{(p_i -eA)^2 }{2 m} + \dfrac{m\Omega^2 x_i^2}{2}}.
	\end{align}
Note that we drop the index $k$ numbering the atoms in $\mathcal{H}_{\rm tm}$ for a moment. As usual, we assume $\mathbf{A}(\mathbf{r}) \approx \mathbf{A} = A_0 (a^\dagger + a)$ in the region where the atoms are located. It is convenient to make the canonical transformation $\tilde{x}_i = - p_i /(m \Omega)$ and $\tilde{p}_i = m \Omega x_i$. This yields 
	\begin{align}
	\mathcal{H}_{\rm tm}^0 =  \sum_{i=1}^{n}{ \left( \dfrac{p_i^2}{2 m} + \dfrac{m\Omega^2 x_i^2}{2} \right) }  + e A_0 \Omega (a^\dagger + a)\sum_{i=1}^n x_i + \dfrac{n e^2 A_0^2}{2m} (a^\dagger + a)^2,
	\end{align}
where we have written $x_i$ and $p_i$ instead of $\tilde{x}_i$ and $\tilde{p}_i$ to keep notation simple. Successively introducing relative and center-of-mass coordinates, $\{x_1,p_1,x_2,p_2\} \rightarrow \{\tilde{x}_1,\tilde{p}_1,X_1,P_1 \}$, $\{X_1,P_1,x_3,p_3 \}\rightarrow \{\tilde{x}_2,\tilde{p}_2,X_2,P_2 \}$, $\ldots$, leads to  
	\begin{align}
	\mathcal{H}_{\rm tm}^0 =  \sum_{i=1}^{n-1}{ \left( \dfrac{\tilde{p}_i^2}{2 \mu_i} + \dfrac{\mu_i\Omega^2 \tilde{x}_i^2}{2} \right) }  
	+ \dfrac{\tilde{P}^2}{2 M} + \dfrac{M \Omega^2 X^2}{2}+
	e A_0 \Omega (a^\dagger + a) n X + \dfrac{n e^2 A_0^2}{2m} (a^\dagger + a)^2.
	\end{align}
Here, $X= X_n = \frac{1}{n} \sum_{j=1}^n x_j$ and $P=P_n= \sum_{j=1}^n p_j$ are the center-of-mass coordinates of all particles in the harmonic oscillator atom, and $M =nm$. The relative coordinates are given by $\tilde{x}_i = (1/i \sum_{j=1}^i x_j) - x_{i+1}$ and $\tilde{p}_i = 1/(i+1) (\sum_{j=1}^i p_j - ip_{j+1})$, and $\mu_i = mi/(i+1)$. Note that the electromagnetic field couples only to the center of mass. With this preliminary work done, one can write the full Hamiltonian as
	\begin{align}
	\mathcal{H}_{\rm tm} = \tilde{\omega} a^\dagger a + \Omega \sum_{k=1}^N{c_k^\dagger c_k} + 
	\tilde{\gamma} \sum_{k=1}^N  ( c_k^\dagger + c_k) (a^\dagger + a) 	+ \Omega \sum_{i = 1}^{N(n-1)} (b_i^\dagger b_i + \dfrac{1}{2}) + C^\prime.
	\end{align} 
The operator $c_k^\dagger$ excites the center-of-mass degree of freedom of the $k$th atom, and the generators for the $N(n-1)$ relative coordinates are denoted by $b_i^\dagger$. We have introduced
	\begin{align}
	\gamma = e A_0 \sqrt{\dfrac{n \Omega }{ 2m }}, \qquad \kappa = nN \dfrac{e^2 A_0^2}{  2m} ,
	\end{align}
and removed the $\kappa$-term by means of $\tilde{\omega} = \sqrt{\omega^2 + 4 \kappa \omega}$ and $\tilde{\gamma} = \sqrt{\omega/\tilde{\omega}} \gamma$ as before ($C^\prime = (\tilde{\omega}-\omega + N\Omega)/2$). The first three terms are again a special case of $\mathcal{H}_{\rm gD}$ with $\Omega_l = \Omega$ and $\tilde{\lambda}_l = \tilde{\gamma}$ for all $l$, and $\mu = N$. Hence, their eigenvalues follow from the roots of the characteristic equation for $\mathcal{H}_{\rm gD}$ (Eq.\ \eqref{eq:chareq}), simplified by the present conditions. They can be explicitly calculated and are the frequencies of the normal modes of field and center-of-mass coordinates. We find $N-1$ eigenfrequencies being equal to $\Omega$ and represent the generators of the corresponding collective excitations also by $b_i^\dagger$. Only two eigenfrequencies $\epsilon_\pm$ are nondegenerate,
	\begin{subequations}
	\begin{align}
	{\mblue 2 \epsilon_{\pm}^2 }&=  \omega^2  + 4 \kappa \omega +  \Omega^2 
	\pm \sqrt{ (\omega^2  +  4\kappa \omega -  \Omega^2 )^2 + 16 N \gamma^2 \omega \Omega}\\
	&{\mblue =  \omega^2  + 4 \kappa \omega +  \Omega^2 
	\pm \sqrt{ (\omega^2  +  4\kappa \omega -  \Omega^2 )^2 + 16 \lambda^2 \omega \Omega} }.
	\end{align}
	\end{subequations}
We have defined $\lambda = \sqrt{N} \gamma$. Since the dipole moment $d$ of the transition from the ground state of an atom to its first excited state is given by {\mblue $d= \bra{n} ex \ket{n-1} = e\sqrt{n/2m\Omega}$}, we can rewrite {\mblue $\lambda = A_0 \Omega d \sqrt{N}$} and {\mblue $\kappa= \lambda^2 / \Omega$}. Denoting the generators of the $\epsilon_\pm$-modes by $a_\pm$, we arrive at 
	\begin{align}
	{\mblue
	\mathcal{H}_{\rm tm} = \epsilon_{\pm} (a^\dagger_{\pm} a_{\pm} + \dfrac{1}{2} )  + \sum_{i=1}^{nN-1}{\Omega (b_i^\dagger b_i + \dfrac{1}{2}) }
	}
	- \dfrac{\omega}{2}.
	\end{align}

\vspace{8mm}
\noindent
{\fontsize{11}{12}\selectfont \textbf{Shift of the resonator frequency due to the $\mathbf{pA}$- and $\mathbf{A}^2$-terms. }}
Consider a system of $N$ mutually noninteracting objects (e.g. atoms) with Hamiltonians
	\begin{align}	
	{\mblue
	H_{\rm mic}^k = \sum_{i=1}^{n_k}{\dfrac{(\mathbf{p}_i^{k})^2}{2m_i^k }} + V_{\rm int}(\mathbf{r}_{1}^{k},\ldots , \mathbf{r}_{n_k}^{k})
	}
	\end{align}
coupled to a field mode of frequency $\omega$. It is described by
	\begin{align}
	{\mblue	
	\mathcal{H}_{\rm mic} = \omega a^\dagger a + \sum_{k=1}^N{ \left( H_{\rm mic}^k + \mathcal{H}_{pA}^k + \mathcal{H}_{A^2}^k \right) }  ,
	}
	\end{align}
where $H_{\rm mic}^k = \sum_{i=1}^{n_k}{(\mathbf{p}_i^{k})^2/2m_i^k } + V_{\rm int}(\mathbf{r}_{1}^{k},\ldots , \mathbf{r}_{n_k}^{k})$, $\mathcal{H}_{pA}^k = - \sum_{i=1}^{n_k}{ q_i^k  \mathbf{A} \mathbf{p}_i^{k}/m_i^k }$, and $\mathcal{H}_{A^2}^k = \sum_{i=1}^{n_k}{  (q_i^k)^2 \mathbf{A}^2/2m_i^k} $. We denote the eigenspectrum of $H_{\rm mic}^k$ by $\{ E_{m_k}	^k,\ket{m_k}_k\}$ and the photon states by $\ket{l}$ and calculate the shifts $\delta \omega _{pA}$ and $\delta \omega _{A^2}$ of the resonator frequency due to $\sum \mathcal{H}_{pA}^k$ and $\sum \mathcal{H}^k_{A^2}$ using the first nonzero terms in a perturbation series for the energy of $\ket{ 0 , \ldots , 0 , l}$. We take $\omega \ll (E_{m_k}^k-E_{0}^k) =: \Omega_{m_k}^k$ for $m_k \neq 0$ and $\mathbf{A} (\mathbf{r}_i^{k}) \approx \mathbf{A} $. With $\mathbf{d}^{k}_{m_k,0} = {}_k\bra{m_k}\sum_{i=1}^{n_k}{q_i^k \mathbf{r}_i^{\, k}} \ket{0}_k$, we find for the $j$th terms $\Delta E_{pA}^j$ and $\Delta E_{A^2}^j$ in the perturbation series for the perturbations $\sum \mathcal{H}_{pA}^k$ and $\sum \mathcal{H}^k_{A^2}$
	\begin{subequations}	
	\begin{align}
	\Delta E_{pA}^1 &= 0 \\
	\Delta E_{pA}^2 &= -  A_0^2 \sum_{k=1}^N{ 
	\sum_{m_k \neq 0}{ \Omega_{m_k}^k | \boldsymbol{\epsilon} \cdot \mathbf{d}^{k}_{m_k,0} |^2 } } 
	\left( \dfrac{(l+1)}{1+ \frac{\omega}{\Omega^k_{m_k}}} +\dfrac{l}{1-\frac{\omega}{\Omega^k_{m_k}}} \right)\\
	& \approx -  A_0^2 \sum_{k=1}^N{ 
	\sum_{m_k \neq 0}{ \Omega_{m_k}^k | \boldsymbol{\epsilon} \cdot \mathbf{d}^{k}_{m_k,0} |^2 } } 
	\left((2l+1) - \left( \dfrac{\omega}{\Omega^k_{m_k}}\right) +(2l+1) \left( \dfrac{\omega}{\Omega^k_{m_k}} \right)^2 \right)\\
	\Delta E_{A^2}^1& = A_0^2 (2l+1)\sum_{k=1}^N \sum_{i=1}^{n_k} \dfrac{(q_i^k)^2}{2m_i^k}
	\end{align}
	\end{subequations}
Therefore,
	\begin{subequations}
	\begin{align}
	\delta \omega_{pA} = & - 2 A_0^2 \sum_{k=1}^N{ 
	\sum_{m_k \neq 0}{ \Omega_{m_k}^k | \boldsymbol{\epsilon} \cdot \mathbf{d}^{k}_{m_k,0} |^2 } } \big( 1 + \frac{\omega^2}{ (\Omega_{m_k}^k) ^2}  \big) \displaybreak[1] \\ 
	\delta \omega_{A^2} = & 2 A_0^2 \sum_{k=1}^{N}{\sum_{i=1}^{n_k}{\dfrac{(q_i^k)^2}{2m_i^k}}} .
	\end{align}
	\end{subequations}
The $\mathbf{pA}$-terms cause a negative and the $\mathbf{A^2}$-terms a positive frequency shift. Note that $\delta \omega_{pA}$ and $\delta \omega_{A^2}$ almost cancel due to the TRK (applied for each $k$). The resulting total frequency shift {\mblue $\delta \omega = \delta \omega_{pA}+\delta \omega_{A^2}$} is suppresed by {\mblue  $\sim (\omega / \Omega_{m_k}^k )^2$} as compared with $\delta \omega_{pA}$ and $\delta \omega_{A^2}$.

\vspace{8mm}
\noindent
{\fontsize{11}{12}\selectfont \textbf{The generalized Dicke Hamiltonian $\mathcal{H}_{\rm gD}$. }} In this section, we derive the Hamiltonian $\mathcal{H}_{\rm gD}$ (Eq.\ \eqref{eq:Hgd_mt})
from $\mathcal{H}_{\rm mic}$ (in the form of Eq.\ \eqref{eq:Hmic_mt}) for $N \rightarrow \infty$ and show how to obtain and evaluate its characteristic equation. 

According to our strategy formulated in the main text, we start from low atomic densities and expand the eigenstates $\ket{\mathcal{E}}$ of $\mathcal{H}_{\rm mic}$ in powers of $A_0 \propto V^{-1/2}$, 
	\begin{align}
	{\mblue \ket{\mathcal{E}} \propto \sum_{s=0}^\infty A_0^s \ket{\mathcal{E}_s}},
	\end{align} 
where $|\mathcal{E}_s \rangle$ stands for a sum over components that each describe $s$ transitions from $\ket{\mathcal{E}_0}$ both in its atomic and in its photonic part and hence has weight $\sim A_0^{s}$ (Fig. \ref{fig:states_gen_no-go}). The corresponding eigenenergies can be written as {\mblue $\mathcal{E} \propto \sum_{ss'} A_0^{s+s'} \langle \mathcal{E}_s | \mathcal{H}_{\rm mic} | \mathcal{E}_{s'} \rangle$}. We are interested only in the low-energy spectrum of $\mathcal{H}_{\rm mic}$. Thus, we assume that the number of atomic excitations {\mblue $\xi = \sum_k \sum_{l>0}| \langle l_k|\mathcal{E}_0\rangle|^2$} and the number of photons $\chi = \bra{\mathcal{E}_0} a^\dagger a \ket{\mathcal{E}_0}$ in the uncoupled eigenstates $\ket{\mathcal{E}_0}$ are small compared to $N$, {\mblue $\xi \ll N$} and $\chi \ll N$. We now calculate $\mathcal{E}$ by dropping all $s+s^\prime \geq 2 $ terms and show that for the low-energy spectrum of $\mathcal{H}_{\rm mic}$ all matrix elements that induce transitions in-between excited atomic states are irrelevant. To that end, we write
	\setcounter{figure}{0}		
	\renewcommand{\thefigure}{A.\arabic{figure}} 
	\begin{figure}
	\centering
	\subfigure[]{\includegraphics[trim = 0pt 0pt 0pt 0pt, width=0.41\textwidth]{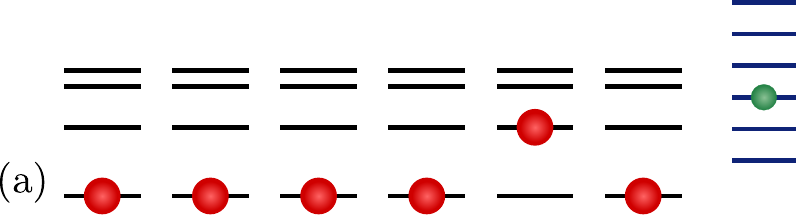}}\hspace{0cm}
	\hspace{1cm}
	\subfigure[]{\includegraphics[width=0.41\textwidth]{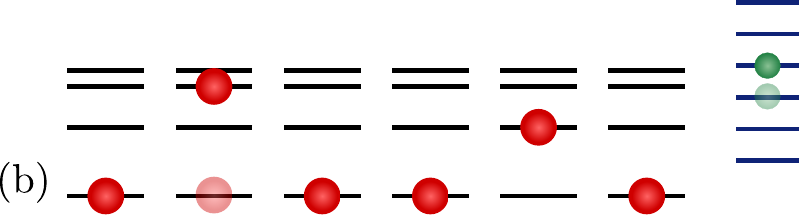}}
	\caption{\label{fig:states_gen_no-go}Situation of the generalized no go theorem. Atomic spectra are drawn black, eigenenergies of the free electromagnetic field blue. 
	(a) Structure of a low-energy state $\ket{\mathcal{E}_0}$ of the uncoupled system. 
	For $N \rightarrow \infty$, the numbers of excited atoms $\xi$ and of photons $\chi$ in $\ket{\mathcal{E}_0}$ 
	are small compared to $N$, $\xi \ll N$ and $\chi \ll N$. 
	(b) Structure of a component of $\mathcal{E}_1$. The coupling has induced one atomic transition and created or annihilated one photon 
	(shown is an excitation of the second atom and the creation of a photon).
	The state $\ket{\mathcal{E}_1}$ is the sum of all such states.
	Their amplitude in the eigenstate of the coupled system is smaller than the amplitude of $\ket{\mathcal{E}_0}$ by a factor $\propto A_0 \propto V^{-1/2}$. 
	In general, $\ket{\mathcal{E}_s}$ represents the sum over all states obtained from $\ket{\mathcal{E}_0}$ via $s$ atomic transitions and $s$ 
	creations or annihilations of a photon. They contribute to the eigenstate of the coupled system by an amplitude $\propto A_0^s$.}
	\end{figure}

	\begin{align}
	\mathcal{H}_{\rm mic} = \tilde{\omega} a^\dagger a +\sum_{k=1}^N
	\sum_{l,l^\prime =0}^{\mu} \Big(
	\Omega_l \delta_{l,l^\prime} \ketbra{l_k}{l_k} 
	+ i A_0\sqrt{\dfrac{\tilde{\omega}}{\omega}} (\Omega_{l^\prime} - \Omega_{l} ) d_{l,l^\prime}(a^\dagger + a ) \ketbra{l_k}{l^{\prime}_k} \Big) +C,
	\end{align}
with $\tilde{\omega} = \sqrt{\omega^2 + 4 \kappa \omega}$ and $C = (\tilde{\omega}-\omega)/2$, and we define
	\begin{subequations}	
	\begin{align}
	H &= \tilde{\omega} a^\dagger a +\sum_{k=1}^N \sum_{l =1}^{\mu} \Omega_l \ketbra{l_k}{l_k} \\
	H_{\rm cpl} &= (a^\dagger + a ) \sum_{k=1}^N \sum_{l =1}^{\mu} A_0\sqrt{\dfrac{\tilde{\omega}}{\omega}} 
	\Omega_{l} \big( id_{0,l} \ketbra{0_k}{l_k} - id_{l,0} \ketbra{l_k}{0_k} \big)\\
	\Delta H & = (a^\dagger + a ) \sum_{k=1}^N \sum_{l > l^\prime \geq 1}^\mu A_0\sqrt{\dfrac{\tilde{\omega}}{\omega}} 
	(\Omega_{l} - \Omega_{l^\prime} )  \big( i d_{l^\prime,l} \ketbra{l_k^\prime}{l_k} - i d_{l,l^\prime} \ketbra{l_k}{l_k^\prime} \big).
	\end{align}
	\end{subequations}
Accordingly,
	\begin{subequations}
	\begin{align}
	\mathcal{E} & \propto \bra{\mathcal{E}} \mathcal{H}_{\rm mic}   \ket{\mathcal{E}} \\
	& \propto \bra{\mathcal{E}_0} H \ket{\mathcal{E}_0} + A_0 \big( \bra{\mathcal{E}_0} H_{\rm cpl} \ket{\mathcal{E}_1} +  \bra{\mathcal{E}_1} H_{\rm cpl} \ket{\mathcal{E}_0} +
	\bra{\mathcal{E}_0} \Delta H \ket{\mathcal{E}_1} +  \bra{\mathcal{E}_1} \Delta H \ket{\mathcal{E}_0}  \big).
	\end{align}
	\end{subequations}
Let us now compare the contributions of $H_{\rm cpl}$ and $\Delta H$ to $\mathcal{E}$. The photonic parts of $H_{\rm cpl}$ and $\Delta H$ are equal and need not be further considered. We write $\langle H_{\rm cpl} \rangle = \bra{\mathcal{E}_0} H_{\rm cpl} \ket{\mathcal{E}_1} +  \bra{\mathcal{E}_1} H_{\rm cpl} \ket{\mathcal{E}_0}$ and $\langle \Delta H \rangle = \bra{\mathcal{E}_0} \Delta H \ket{\mathcal{E}_1} +  \bra{\mathcal{E}_1} \Delta H \ket{\mathcal{E}_0} $, and we find
	\begin{align}
	\dfrac{\langle \Delta H \rangle }{\langle H_{\rm cpl} \rangle } = \dfrac{ \sum_{l> l^\prime \geq 1}^\mu (\Omega_{l}- \Omega_{l^\prime})  {\rm Im} \big[ d_{l^\prime,l} \sum_{k=1}^N \big(\braket{\mathcal{E}_0}{l_k^\prime}\braket{l_k}{\mathcal{E}_1} +  \braket{\mathcal{E}_1}{l_k^\prime}\braket{l_k}{\mathcal{E}_0}\big) \big]}{\sum_{l=1}^\mu \Omega_l {\rm Im} \big[ d_{0,l} \sum_{k=1}^N \big( \braket{\mathcal{E}_0}{0_k}\braket{l_k}{\mathcal{E}_1} +  \braket{\mathcal{E}_1}{0_k}\braket{l_k}{\mathcal{E}_0} \big)\big] }
	\end{align}
Since $N \rightarrow \infty$, the number of nonzero terms in the $k$-sums is decisive. For given $l,l^\prime$, the sum over $k$ in the numerator has at most $\xi$ nonzero terms, whereas the sum over $k$ in the denominator has at least $N-\xi$ nonzero terms. Hence, we drop $\Delta H$, which represents the matrix elements of $\mathcal{H}_{\rm mic}$ connecting the excited states of an atom, and keep $H_{\rm cpl}$ as the relevant coupling part of $\mathcal{H}_{\rm mic}$. We reintroduce the $\kappa$-term and call the resulting Hamiltonian \emph{generalized Dicke Hamiltonian}~$\mathcal{H}_{\rm gD}$,
	\begin{align}\label{eq:H_gDI}
	\mathcal{H}_{\rm gD} =\omega a^\dagger a + \kappa (a^\dagger + a)^2 +
	\sum_{k=1}^N
	\sum_{l =1}^{\mu}  \left(
	\Omega_l  \ketbra{l_k}{l_k} 
	-  \Omega_l A_0 (a^\dagger + a) \big( i d_{l,0} \ketbra{l_k}{0_k}  + {\rm H.c.}\big) \right).
	\end{align}
It has the same low-energy spectrum as $\mathcal{H}_{\rm mic}$. Paralleling our treatment of $\mathcal{H}_{\rm D}$, we introduce
	\begin{align}
	a_{k,l}^\dagger = \ketbra{l_k}{0_k}, \qquad  b_{q_j,l}^\dagger = \dfrac{1}{\sqrt{N}} \sum_{k=1}^N e^{i q_j k} \ketbra{l_k}{0_k},
	\end{align}
where $q_j = 2\pi (j/N)$ and $j \in \{0,1,\ldots,N-1\}$ as before. With $\sum_{k=1}^{N} a_{k,l}^\dagger a_{k,l} = \sum_{j=0}^{N-1} b_{q_j,l}^\dagger b_{q_j,l}  $, Eq.\ \eqref{eq:H_gDI} becomes
	\begin{align}
	\mathcal{H}_{\rm gD} =\omega a^\dagger a + \kappa (a^\dagger + a)^2 +
	\sum_{l =1}^{\mu}  \Big(
	\Omega_l  \sum_{j=0}^{N-1}  b_{q_j,l}^\dagger b_{q_j,l}
	-  A_0 \Omega_l \sqrt{N}  (a^\dagger + a)\big( i d_{l,0} b_{q_0,l}^\dagger  + {\rm H.c.}\big) \Big).
	\end{align}
The operators $b_{q_j,l}$ are bosonic in the limit of dilute excitations ($\xi \ll N$)\cite{Hopfield1958_app}. Again, we drop the energy of the dark modes, write $b_l$ instead of $b_{q_0,l}$, define {\mblue $\lambda = A_0 \Omega_l |d_{0l}| \sqrt{N}$}, and remove the $\kappa$-term by substituting {\mblue $\omega \rightarrow \tilde{\omega} = \sqrt{\omega^2 + 4 \kappa\omega}$} and {\mblue $\lambda_l \rightarrow \tilde{\lambda_l} = \sqrt{\omega/\tilde{\omega}} \lambda_l$} and adding $C= (\tilde{\omega}-\omega)/2$. This gives
	\begin{align}\label{eq:H_gD}
	{\mblue
	\mathcal{H}_{\rm gD} = \tilde{\omega} a^\dagger a + 
	\sum_{l=1}^\mu{  \Omega_l b_l^\dagger b_l }+ \sum_{l=1}^{\mu}{\tilde{\lambda}_l(b_l^\dagger + b_l)(a^\dagger +a) } 
	}	+ C.
	\end{align}
In order to find the eigenfrequencies of $\mathcal{H}_{\rm gD}$, we introduce canonical coordinates by means of
	\begin{align}
	x=\dfrac{1}{\sqrt{2  \tilde{\omega}}} (a^\dagger + a), \quad p_{x} = i \sqrt{ \dfrac{ \tilde{\omega}}{ 2 }} (a^\dagger - a), 
 	\quad y_l =\dfrac{1}{\sqrt{2 \Omega_l}} (b_l^\dagger + b_l), \quad p_{l} = i \sqrt{\dfrac{ \Omega_l}{ 2 }} (b_l^\dagger - b_l),
	\end{align}
and define $\mathbf{X}^{\rm T} = (x,y_1,\ldots,y_\mu)$, $\mathbf{P}^{\rm T}=(p_x,p_1,\ldots, p_\mu)$, and $g_l= 2 \tilde{\lambda}_l \sqrt{\tilde{\omega} \Omega_l}$. This yields
	\begin{align}
	\mathcal{H}_{\rm gD} = \dfrac{\mathbf{P}^{\rm T} \mathbf{P}}{2}   +  \dfrac{1}{2} \mathbf{X}^{\rm T} \underline{\Omega}^2\mathbf{X}  - \dfrac{1}{2}\big(\omega +\sum_{l=1}^\mu \Omega_l)
	\end{align}
where
	\begin{align}
	\underline{\Omega}^2 = 
	\left(
	\begin{matrix}
	\tilde{\omega}^2 & g_1  & \cdots  &g_\mu \\
	g_1  &  \Omega_1^2 &  & \\
	\vdots & & \ddots &  \\
	g_\mu  &  & &  \Omega_\mu^2 
	\end{matrix}
	\right)
	\end{align}
The orthogonal matrix $G$ that diagonalizes $\underline{\Omega}^2$ induces a point transformation to the normal modes $\tilde{\mathbf{X}} = G \mathbf{X}$ and $\tilde{\mathbf{P}} = G \mathbf{P}$. The eigenvalues $\epsilon_i^2$ of $\underline{\Omega}^2$ are the squared eigenfrequencies of the system. They solve the characteristc equation
	\begin{subequations}
	\begin{align}
	{\mblue  0 
	}&
	= \Big( \prod_{l^\prime =1}^{\mu}{(\Omega_{l^\prime}^2-\epsilon^2)}\Big) \Big( (\tilde{\omega}^2 -\epsilon^2) -  
	\sum_{l=1}^{\mu }{\dfrac{g_l^2  }{ \Omega_l^2 -\epsilon^2}  }\Big) \\
	& {\mblue =
	\Big( \prod_{l^\prime =1}^{\mu}{(\Omega_{l^\prime}^2 - \epsilon^2)} \Big)   \Big( (\tilde{\omega}^2 -\epsilon^2) 
	 - 4 \tilde{\omega}\sum_{l =1}^{\mu}{\dfrac{ \Omega_{l} \tilde{\lambda}_{l}^2 }{\Omega_{l}^2 - \epsilon^2}} \Big)
	}.
	\label{eq:chareq}
	\end{align}	
	\end{subequations}
None of them can be zero since this would imply 
	\begin{align}\label{eq:contradiction}	
	{\mblue	
	\dfrac{\omega}{4N A_0^2} =  \sum_{l=1}^{\mu}{\Omega_l} |d_{0l}|^2 -\sum_{i=1}^{n}{\dfrac{q_i^2}{2m_i}}
	}.
	\end{align}
We have used {\mblue $\tilde{\omega} = \sqrt{\omega^2 + 4 \kappa \omega }$}, {\mblue $\tilde{\lambda}_l = \sqrt{\omega/\tilde{\omega}} \lambda_l$}, {\mblue $\lambda_l = A_0 \Omega_l |d_{0l}| \sqrt{N} $}, and $\kappa = N A_0^2 \sum_{i=1}^n q_i^2/2m_i$. However, the left side of Eq.~\eqref{eq:contradiction} is positive, whereas its right side is negative according to the TRK for $H_{\rm mic}^0$ (Eqs.\ \eqref{eq:TRK} or Eq.\ \eqref{eq:TRKcircuit} of the main text).

\end{document}